\documentclass[prl,twocolumn]{revtex4}
\usepackage{amsmath}
\usepackage{amssymb}
\usepackage{graphicx}
\begin{document}

\title{"0.7 anomaly" and magnetic impurity formation in quantum point contacts}

\author{S. Ihnatsenka \& I. V. Zozoulenko}
\affiliation{Solid State Electronics, Department of Science and Technology (ITN), Link\"{o}ping University, 60174 Norrk\"{o}ping, Sweden}

\maketitle

\textbf{The origin of the 0.7-conductance anomaly in quantum point contacts (QPCs) has
been a subject of lively discussions since its discovery more than 10 years ago \cite{1}. In
a recent letter, based on spin density functional theory (DFT), Rejec and Meir
explained the origin of the 0.7-anomaly as being due to the formation of a magnetic
impurity in the QPCs \cite{2}. They did not, however, perform transport conductance
calculations so the central question whether the 0.7-anomaly is indeed related to
the formation of a magnetic impurity has remained unanswered. In this
communication, we perform spin DFT \textit{transport} calculations for the structures
considered by Rejec and Meir. While we recover the findings reported by Rejec
and Meir concerning the formation of localized spin-degenerate quasi-bound
states, our transport calculations do not contain the 0.7 anomaly in the
conductance. We suggest that the inability of the DFT to reproduce quantitatively
the 0.7 anomaly may be due to the uncorrected self-interaction errors in the DFT
calculation of electronic transport for the case when localization of charge is
expected to occur, so that the magnetic impurity formation may be an artefact of
DFT due to the spurious self-interaction.
}

We consider quantum point contacts defined by a split-gate in GaAs
heterostructure similar to those studied by Rejec and Meir, see Fig. 1. We utilize the
same Hamiltonian based on the spin DFT approximation (a detailed description of the
Hamiltonian can be found in Refs. \cite{3,4,5}). We calculate the
scattering solutions of the Schrodinger equation in the open system using the selfconsistent
Green's function technique where the whole device, including the semiinfinitive
leads, is treated on the same footing, i.e. the electron-electron interaction is
accounted for both in the leads and in the QPC region within the spin DFT
approximation. The detailed description of our method is given in Ref. \cite{5}. Note that
similar spin DFT conductance calculations reproduce quantitatively the measured spinresolved
magnetoconductance of quantum wires in the integer quantum Hall regime \cite{6,7}.

Figures 1 a, b show the spin resolved electron densities, the local density of states
and the potentials for spin-up and spin-down electrons in the QPC of the lithographic
length of 250 nm. These results agree very well with corresponding findings reported by
Rejec and Meir \cite{2}. (Note that spin DFT calculations predicting spin polarization in the
QPC were reported in Refs. \cite{8,9,10}). The calculated conductance is shown in Fig. 1 c.
Close to the pinch off a spin-degeneracy of the spin-up and spin-down conductance
channels is lifted and the total conductance shows a broad feature peaked at $\sim 0.5\times2e^{2}/h$.
A similar feature is also present in the range of the gate voltages where a second step in
the conductance develops. The calculated conductance clearly does not reproduce the
0.7 anomaly observed in almost all QPCs of various geometries. (We stress that the
results presented here are generic; we studied QPCs with lengths in the range 40-400
nm and electron densities in the leads in the range $10^{15}$ m$^{-2}$ - 4$\times10^{15}$ m$^{-2}$, with very
similar results).

Why do DFT calculations fail to reproduce the 0.7 anomaly? Formation of the
magnetic moment implies that electrons are trapped in the weakly coupled quasi-bound
states in the center of the QPC \cite{2}. It has been recently realized that a standard DFT
approach in the case of the weak coupling suffers from the spurious interaction of the
electron with its exchange and correlation potential \cite{11,12,13,14,15}. This, for example, causes an
orders-of-magnitude discrepancy between the calculated and measured currents through
organic molecules coupled to metallic contacts \cite{11,12,13,14,15} (This is in contrast to the case of
metallic atomic-scale wires for which the DFT-based calculations are in excellent
agreement with the experiment \cite{11,12}). Several correction schemes have been recently
suggested and implemented, restoring the agreement between the DFT calculations and
the experiment for the case of the weak coupling \cite{11,12,13,14,15}. A spurious self-interaction might
well be the reason for the failure of the standard DFT for the case of the QPC and an
accurate description of the 0.7 anomaly might require correction schemes eliminating
the above errors.

\begin{figure*}[tb]
\includegraphics[scale=1.4]{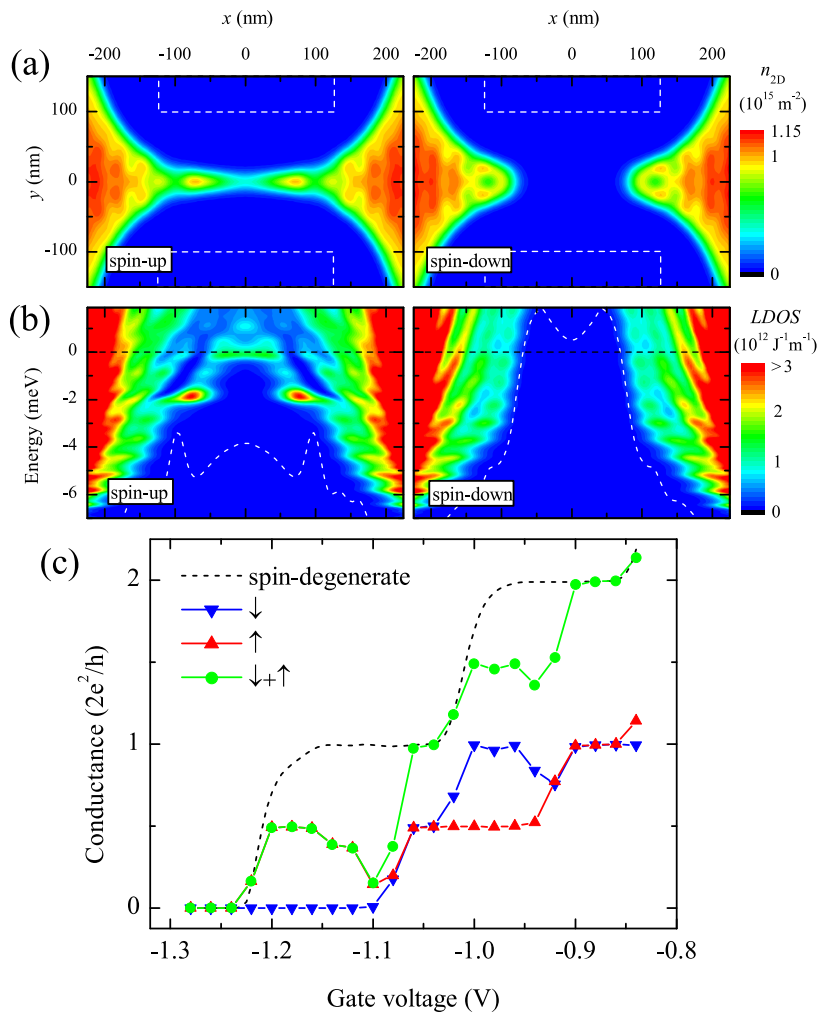}
\caption{Figure 1 $\mid$ The conductance and the electron densities in the QPC. (a) The
spin-up (left) and spin-down (right) electron densities in the quantum point
contact for the gate voltage $V_{g}=-1.2$ V and (b) the corresponding local density
of states (LDOS). The dotted lines indicate the self-consistent Kohn-Sham
potential in the center of the wire (along $y=0$). (c) The spin resolved and the
total conductance of the QPC ($\uparrow$, $\downarrow$ stand for the spin-up and spin-down
electrons). The dotted line corresponds to the spin-unpolarized
conductance. The spin polarized conductance was obtained by applying a small
magnetic field in the initial iterations of the self-consisted procedure. The
parameters of the structure are as follows. The QPC is defined in the infinite
quantum wire of the width of 500 nm. The lithographic position of the gates
defining the QPC is indicated by the dashed lines in (a). The distance between
the gates is 200 nm, and the gate length is 250 nm. The electrons are situated
50 nm below the surface of the GaAs heterostructure. The donor layer of the
width of 36 nm and concentration of $7.2\times10^{23}$ m$^{-3}$ is situated 10 nm below the
surface. The electron density in the quantum wire far away from the QPC is
$2.5\times10^{15}$ m$^{-2}$ (which corresponds to 18 propagating modes in the leads). (A
detailed description of the model for the heterostructure and the confining
potential can be found in Refs. \cite{3,4,5}). The calculations are performed for $T=200$
mK.}
\end{figure*}


\begin{thebibliography}{15}
\bibitem{1} Thomas, K. J., Nicholls, J. T., Simmons, M. Y., Pepper, M., Mace, D. R., \& Ritchie, D. A., \textsl{Phys. Rev. Lett.} \textbf{77}, 135-138 (1996).
\bibitem{2} Rejec, T., \& Meir, Y. \textsl{Nature} \textbf{442}, 900-903 (2006).
\bibitem{3} Ihnatsenka, S., \& Zozoulenko, I. V. \textit{Phys. Rev. B} \textbf{73}, 075331 1-14 (2006).
\bibitem{4} Ihnatsenka, S., \& Zozoulenko, I. V. \textit{Phys. Rev. B} \textbf{74}, 201303(R) 1-4 (2006).
\bibitem{5} Ihnatsenka, S., Zozoulenko, I. V, \& Willander, M. cond-mat/0701107 (submitted to Phys. Rev. B).
\bibitem{6} Ihnatsenka, S. \& Zozoulenko, I. V. to be submitted to Phys. Rev. B.
\bibitem{7} Radu, I. P., Miller, J. B., Amasha, S., Levenson-Falk, E., Zumbuhl, D. M., Kastner, M. A., Marcus, C. M., Pfeiffer L. N., \& West, K. W., to be submitted to Phys. Rev. B
\bibitem{8} Berggren, K.-F. \& Yakimenko, I. I., \textit{Phys. Rev. B} \textbf{66}, 085323 1-7 (2002).
\bibitem{9} Hirose, K., Meir, Y. \& Wingreen, N. S., \textit{Phys. Rev. Lett.} \textbf{90}, 026804 1-4 (2003).
\bibitem{10} Havu, P., Puska, M. J., Nieminen, R. M. \& Havu, V., \textit{Phys. Rev. B} \textbf{70}, 233308 1-4 (2004).
\bibitem{11} Evers, F., Weigend, F., \& Koentopp, M. \textit{Phys. Rev. B} \textbf{69}, 235411 1-9 (2004).
\bibitem{12} Sai, N., Zwolak, M., Vignale, G., \& Di Ventra, M. \textit{Phys. Rev. Lett.} \textbf{94}, 186810 1-4 (2005).
\bibitem{13} Toher, C., Filippetti, A., S. Sanvito, S., \& Burke, K. \textit{Phys. Rev. Lett.} \textbf{95}, 146402 1-4 (2005).
\bibitem{14} Palacios, J. J. \textit{Phys. Rev. B} \textbf{72}, 125424 1-6 (2005).
\bibitem{15} Muralidharan, B., Ghosh, A. W., \& Datta, S. \textit{Phys. Rev. B} \textbf{73}, 155410 (2006).

\end{thebibliography}
\end{document}